\documentclass[aps,prb,twocolumn,groupedaddress,citeautoscript,dvipdfmx]{revtex4}
\usepackage{bm,amsmath,amssymb,graphicx}
\begin{document} 
\title{Gapless surface states in a three-dimensional Chalker-Coddington type network model}
\author{Tetsuyuki Ochiai}
\affiliation{Photonic Materials Unit, National Institute for Materials Science, Tsukuba 305-0044, Japan}
\date{\today}

\begin{abstract}
We present the emergence of gapless surface states in a three-dimensional Chalker-Coddington type network model with spatial periodicity. The model consists of a ring network placed on every face of the cubic unit cells in the simple cubic lattice. The scattering among ring-propagating modes in the adjacent rings is described by the S-matrices, which control possible symmetries of the system.   
The model maps to a Floquet-Bloch system, and the quasienergy spectrum can exhibit a gapped bulk band structure and gapless surface states.  
Symmetry properties of the system and robustness of the gapless surface states  are explored in comparison to topological crystalline insulator.   
We also discuss other crystal structures, a gauge symmetry, and a possible optical realization of the network model.  
\end{abstract}

\maketitle

\section{Introduction}

Recently, much attention has been paid to two-dimensional (2d) ring-resonator arrays as a platform of synthetic gauge fields for photons \cite{hafezi2013imaging}. An effective magnetic fields can be implemented in such a system through optical path phases, which act as the Aharonov-Bohm phases for photons. As a result, a nontrivial topology of bulk modes and a chiral edge mode of photons can be realized in such a system.  
Provided that the ring-resonator system does not contain any non-reciprocal elements, the effective magnetic field is spin-dependent, preserving the  time-reversal symmetry. 
 Here, the term ``spin'' is referred to as clockwise or counter-clockwise mode in a ring resonator.  
The effective magnetic field in the time-reversal-invariant system is reminiscent of the spin-orbit interaction in the Kane-Mele model \cite{kane2005qsh} of topological insulator (TI).

Interestingly, even if the effective (spin-dependent) magnetic field is zero, the ring-resonator array can exhibit nontrivial topological phases \cite{PhysRevLett.110.203904}. The essence is the directional coupling among the ring resonators and the mapping to a Floquet-Bloch system \cite{PhysRevB.89.075113}. By the directional coupling, the spin degrees of freedom are decoupled, and the time-reversal symmetry appears to be broken for a given decoupled spin sector. The system is like the (2+1)-dimensional massive Dirac system, which exhibits the quantum Hall effect under vanishing magnetic field \cite{semenoff1984cms,haldane1988mqh}. 
By the mapping to Floquet-Bloch, the system acquires a nontrivial topology via wrapped quasienergy \cite{PhysRevB.82.235114}.   
In fact, the ring-resonator array can be viewed as an optical realization of the Chalker-Coddington network model \cite{0022-3719-21-14-008}. The model on a 2d lattice without disorder can be mapped to a Floquet-Bloch system, which has the quasienergy spectrum as a function of Bloch momentum \cite{PhysRevB.59.15836}. 
A  mapping to the (2+1)-dimensional massive Dirac system is also available \cite{PhysRevB.54.8708}.  
A part of the topological phases corresponds to the Chern insulator, whereas the rest is the anomalous Floquet insulator with vanishing Chern number. In this way, the 2d ring-resonator lattice can be a new type of photonic (Floquet) TIs.

So far, various proposals have been made to construct photonic TIs. They include  the constructions via the magnetoelectric effect \cite{khanikaev2013photonic,chen2014experimental,ochiai2015time},  helix waveguide arrays \cite{rechtsman2013photonic}, ring-resonator arrays \cite{hafezi2013imaging,PhysRevLett.110.203904}, and hexagonal arrangements of cylinder arrays \cite{PhysRevLett.114.223901}.   
However, to the  best of  my knowledge, a limited number of  proposals has been made for three-dimensional (3d) photonic topological (crystalline) insulators \cite{PhysRevB.84.195126,lu2015three}.   
Therefore, it is still challenging to realize 3d photonic TIs.  
We expect that the network model is an interesting approach to 3d TIs, as 
the model does not rely on any unconventional physical effects and assumptions.

An important aspect of the network model is that it is essentially a bosonic system.  
Although the spin degrees of freedom exist as a pair of clockwise and counter-clockwise modes, the time-reversal symmetry ${\cal T}$ in the network model is bosonic, satisfying ${\cal T}^2=1$. Therefore, the Kramers degeneracy is absent. As a result, a spin-flip process destroys the gapless edge states in the 2d network model \cite{PhysRevB.89.075113}. We should note that intrinsic one-half spin with the Kramers degeneracy can be introduced theoretically in the network model \cite{PhysRevB.76.075301,1367-2630-12-6-065005}. 
However, its optical realization will be difficult. We may need the electromagnetic duality in the optical network \cite{ochiai2015time}.  
To overcome this circumstance without the Kramers degeneracy, crystal symmetries in a bosonic 3d network system may 
protect possible gapless surface states as in topological crystalline insulator (TCI) \cite{fu2011topological}.

In this paper, we propose a 3d Chalker-Coddington type network model 
and analyze its fundamental properties. Eigenvalue equations  for the bulk and surface modes are derived in a Floquet-Bloch form.  Possible symmetry of the system is encoded in the S-matrices, basic ingredients of the network model.  
 Respecting the full symmetry of $O_h^{1}$ on the simple cubic lattice, we numerically study the bulk and surface quasienergy band structures and find gapless surface states in the bulk band gaps. The robustness of the gapless surface states is discussed in connection with the TCI \cite{fu2011topological}.  A possible optical realization of the 3d network model via ring and spherical  resonator, other crystal structure than the simple cubic lattice, and an implementation of the U(1) gauge symmetry are also discussed.

Other types of 3d network models have been proposed and studied in detail within the context of the Anderson localization in layered quantum-hall and quantum-spin-hall systems \cite{PhysRevLett.75.4496,klesse1999modeling,PhysRevB.89.155315}. In contrast to these models, our model is truly three-dimensional, not simply a layered 2d system. We consider that this property with high symmetry of $O_h^1$ has profound effects in the nontrivial band structure.

This paper is organized as follows.  In Sec. II, we define the 3d network model. Eigenvalue equations are derived for bulk and surface modes. We discuss symmetry properties of the system in Sec. III. Bulk and surface band structures are presented for various inputs, and gapless surface states are found in Sec. IV. 
Their robustness is discussed.   
Summary and outlook are given in Sec. V, discussing other crystal structures, a gauge symmetry, and an optical realization.

\section{Model}

\begin{figure}
\includegraphics*[width=0.45\textwidth]{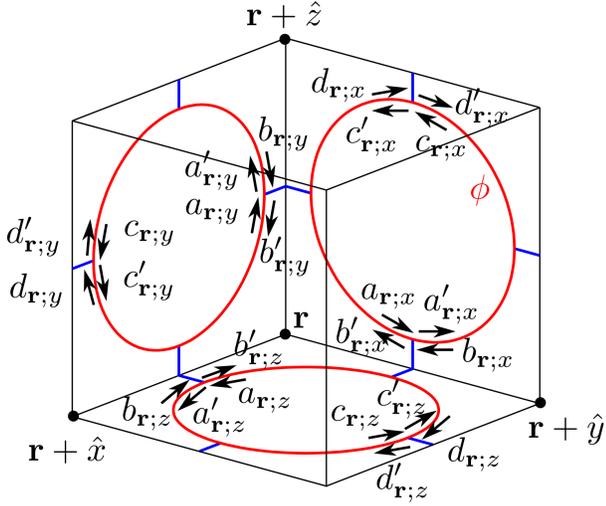} 
\caption{\label{Fig_unit} (Color online) Unit cell of the three-dimensional Chalker-Coddington type network model. It consists of identical rings  placed on every face of the cubic unit cell. For visibility, rings in the front two and top faces are omitted. Adjacent rings are connected by legs (denoted by blue). Modes are propagating in the rings, and are scattered at the legs.   
}
\end{figure}

Here, we present a 3d Chalker-Coddington type network model. 
The system under study is  shown in Fig. \ref{Fig_unit}. 
It consists of identical rings  placed on every face of the cubic unit cells in the simple cubic lattice.  
In each ring, clockwise and counter-clockwise modes are introduced.  
A scattering takes place among adjacent rings through the legs (indicated by blue lines in Fig. \ref{Fig_unit}) placed  at the hinges of the unit cells.  
The mode amplitudes $\alpha_{{\bm r};j}$ and $\alpha'_{{\bm r};j}$ ($\alpha=a,b,c,d$ and $j=x,y,z$) acquire phase $\phi$ for every one-quarter-ring propagation between the legs.

\begin{figure}
\includegraphics*[width=0.45\textwidth]{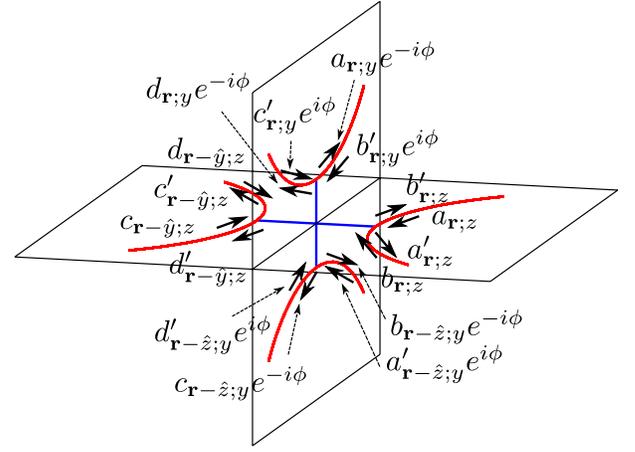}
\caption{\label{Fig_Smatrix} (Color online) Amplitudes relevant to the S-matrix in the $x$ direction. It relates  eight incoming mode amplitudes to eight outgoing mode amplitudes connected by the legs.  The other S-matrices ($S_y$ and $S_z$) are  defined by the cyclic rotations of $x$, $y$, and $z$ indices. }
\end{figure}

The S-matrices are defined among four adjacent rings connected by the legs as shown in Fig. \ref{Fig_Smatrix}. 
Since the S-matrices are associated with the hinges, there are three S-matrices , $S_{{\rm r};x}$, $S_{{\bm r};y}$, and $S_{{\bm r};z}$ per unit cell.   They relate eight incoming-mode amplitudes  to eight outgoing-mode amplitudes as 
\begin{align}
& \left(\begin{array}{l}
a_{{\bm r};z}'\\
b_{{\bm r};z}'\\
c_{{\bm r}-\hat{y};z}'\\
d_{{\bm r}-\hat{y};z}'\\
a_{{\bm r};y}e^{-i\phi}\\
d_{{\bm r};y}e^{-i\phi}\\
b_{{\bm r}-\hat{z};y}e^{-i\phi}\\
c_{{\bm r}-\hat{z};y}e^{-i\phi}
\end{array}\right)=S_{{\bm r};x}
\left(\begin{array}{l}
a_{{\bm r};z}\\
b_{{\bm r};z}\\
c_{{\bm r}-\hat{y};z}\\
d_{{\bm r}-\hat{y};z}\\
c_{{\bm r};y}'e^{i\phi}\\
b_{{\bm r};y}'e^{i\phi}\\
d_{{\bm r}-\hat{z};y}'e^{i\phi}\\
a_{{\bm r}-\hat{z};y}'e^{i\phi}
\end{array}\right),\\
& \left(\begin{array}{l}
a_{{\bm r};x}'\\
b_{{\bm r};x}'\\
c_{{\bm r}-\hat{z};x}'\\
d_{{\bm r}-\hat{z};x}'\\
a_{{\bm r};z}e^{-i\phi}\\
d_{{\bm r};z}e^{-i\phi}\\
b_{{\bm r}-\hat{x};z}e^{-i\phi}\\
c_{{\bm r}-\hat{x};z}e^{-i\phi}
\end{array}\right)=S_{{\bm r};y}
\left(\begin{array}{l}
a_{{\bm r};x}\\
b_{{\bm r};x}\\
c_{{\bm r}-\hat{z};x}\\
d_{{\bm r}-\hat{z};x}\\
c_{{\bm r};z}'e^{i\phi}\\
b_{{\bm r};z}'e^{i\phi}\\
d_{{\bm r}-\hat{x};z}'e^{i\phi}\\
a_{{\bm r}-\hat{x};z}'e^{i\phi}
\end{array}\right),\\
& \left(\begin{array}{l}
a_{{\bm r};y}'\\
b_{{\bm r};y}'\\
c_{{\bm r}-\hat{x};y}'\\
d_{{\bm r}-\hat{x};y}'\\
a_{{\bm r};x}e^{-i\phi}\\
d_{{\bm r};x}e^{-i\phi}\\
b_{{\bm r}-\hat{y};x}e^{-i\phi}\\
c_{{\bm r}-\hat{y};x}e^{-i\phi}
\end{array}\right)=S_{{\bm r};z}
\left(\begin{array}{l}
a_{{\bm r};y}\\
b_{{\bm r};y}\\
c_{{\bm r}-\hat{x};y}\\
d_{{\bm r}-\hat{x};y}\\
c_{{\bm r};x}'e^{i\phi}\\
b_{{\bm r};x}'e^{i\phi}\\
d_{{\bm r}-\hat{y};x}'e^{i\phi}\\
a_{{\bm r}-\hat{y};x}'e^{i\phi}
\end{array}\right),
\end{align}
Suppose that the S-matrices are common in all the unit cells, namely, $S_{{\bm r};j}=S_j$, we have the spatial periodicity of the simple cubic lattice. Then, the Bloch theorem can be applied. We thus have, for instance, $c_{{\bm r}-\hat{y};z}=\exp(-ik_y)c_{{\bm r};z}$ for the mode amplitude. Here, the lattice constant is taken to be unity.

After some algebra, the above equations of the S-matrices are cast into the
eigenvalue equation $U_{\bm k}A =\exp(-i\phi)A$ with  
\begin{widetext} 
\begin{align}
&U_{\bm k}=\left(\begin{array}{cccccc}
0 & K_y^{-1}S_z^{-+}P_x & 0 & K_y^{-1}S_z^{--}M_y & 0 & 0 \\
0 & 0 & K_z^{-1}S_x^{-+}P_y & 0 & K_z^{-1}S_x^{--}M_z & 0 \\
K_x^{-1}S_y^{-+}P_z & 0 & 0 & 0 & 0 & K_x^{-1}S_y^{--}M_x \\
P_z^{-1}S_y^{++}P_z & 0 & 0 & 0 & 0 & P_z^{-1}S_y^{+-}M_x \\
0 & P_x^{-1}S_z^{++}P_x & 0 & P_x^{-1}S_z^{+-}M_y & 0 & 0 \\
0 & 0 & P_y^{-1}S_x^{++}P_y & 0 & P_y^{-1}S_x^{+-}M_z & 0
\end{array}\right),\quad 
A=\left(\begin{array}{c}
A_x\\
A_y\\
A_z\\
\tilde{A}_x'\\
\tilde{A}_y'\\
\tilde{A}_z'
\end{array}\right),\\
&P_j=\left(\begin{array}{cccc}
1 & 0 & 0 & 0\\
0 & 1 & 0 & 0\\
0 & 0 & e^{-ik_j} & 0\\
0 & 0 & 0 & e^{-ik_j} 
\end{array}\right),\quad 
M_j=\left(\begin{array}{cccc}
0 & 0 & 1 & 0\\
0 & 1 & 0 & 0\\
0 & 0 & 0 & e^{-ik_j}\\
e^{-ik_j} & 0 & 0 & 0
\end{array}\right),\quad  
K_j=\left(\begin{array}{cccc}
1 & 0 & 0 & 0\\
0 & 0 & 0 & 1\\
0 & e^{-ik_j} & 0 & 0\\
0 & 0 & e^{-ik_j} & 0
\end{array}\right),\\
&A_j=(a_j,b_j,c_j,d_j)^t,\quad \tilde{A}_j'=e^{i\phi}(a_j',b_j',c_j',d_j')^t,
\end{align}
\end{widetext}
where the $8\times 8$ S-matrix $S_j$ is divided into $4\times 4$ block matrices $S_j^{\pm\pm}$ as 
\begin{align}
S_j=\left(\begin{array}{cc}
S_j^{++} & S_j^{+-}\\
S_j^{-+} & S_j^{--}\\
\end{array}\right). 
\end{align}
The position ${\bm r}$ dependence of the mode amplitudes is omitted. 
If the S-matrix is unitary $S_j^\dagger S_j=1$, matrix $U_{\bm k}$ is shown to be unitary, and thus the eigenvalues of  $\phi$ are real.

In an optics viewpoint, the ring is locally a straight waveguide, and a waveguide mode is characterized by a dispersion relation $\beta(\omega)$ for the propagation constant $\beta$ and angular frequency $\omega$. The dispersion relation is usually monotonic with respect to $\omega$.  
The propagation phase $\phi$ acquired for one quarter of the ring corresponds to 
$\beta(\omega)R\pi/2$, being $R$ the radius of the ring. Therefore, $\phi$ can be viewed as energy, though $\hbar\omega$ should be called it.

The eigenvalue equation has a similarity to that in a Floquet-Bloch system. 
Namely, $U_{\bm k}$ corresponds to the time-translation operator for one period of a Floquet hamiltonian, and  $\phi$ is defined modulo $2\pi$. We thus call $\phi$ the quasienergy. The quasienergy spectrum has the well-defined band structure, because eigenvalues of $\phi$ are real provided that the S-matrices are unitary.

We are also interested in the surface modes in a slab geometry of the network model. For simplicity, we consider here the surfaces normal to the $z$ direction. 
In this  case, we cannot employ the Bloch theorem for the $z$ direction. Instead, we must hold $z$ coordinate index $n\in Z$ for mode amplitudes as $a_{n;j}$ etc. The eigenvalue equation to be solved becomes  
\begin{align}
& K_y^{-1}S_z^{-+}P_xA_{ny} + K_y^{-1}S_z^{--}M_y\tilde{A}_{nx}' = e^{-i\phi}A_{nx},\\      
& K_0^{-1}S_x^{-+}P_yA_{nz} + K_0^{-1}S_x^{--}M_0\tilde{B}_{ny}' = e^{-i\phi}B_{ny},\\ 
& K_x^{-1}S_y^{-+}B_{nx} + K_x^{-1}S_y^{--}M_x\tilde{A}_{nz}' = e^{-i\phi}A_{nx},\\
&S_y^{++}B_{nx} + S_y^{+-}M_x\tilde{A}_{nz}' = e^{-i\phi}\tilde{B}_{nx}',\\
&P_x^{-1}S_z^{++}P_xA_{ny} + P_x^{-1}S_z^{+-}M_y\tilde{A}_{nx}' = e^{-i\phi}\tilde{A}_{nx}',\\
&P_y^{-1}S_x^{++}P_yA_{nz} + P_y^{-1}S_x^{+-}M_0\tilde{B}_{ny}' = e^{-i\phi}\tilde{B}_{ny}',\\ 
&A_{nj}=(a_{n;j},b_{n;j},c_{n;j},d_{n;j})^t,\\
&\tilde{A}_{nj}'=e^{i\phi}(a_{n;j}',b_{n;j}',c_{n;j}',d_{n;j}')^t,\\
&B_{nx}=(a_{n;x},b_{n;x},c_{n-1;x},d_{n-1;x}),\\  
&B_{ny}=(a_{n;y},b_{n-1;y},c_{n-1;y},d_{n;y}),\\
&\tilde{B}_{nx}'=e^{i\phi}(a_{n;x}',b_{n;x}',c_{n-1;x}',d_{n-1;x}'),\\
&\tilde{B}_{ny}'=e^{i\phi}(a_{n-1;y}',b_{n;y}',c_{n;y}',d_{n-1;y}'),\\
&M_0=\left(\begin{array}{cccc}
0 & 0 & 1 & 0\\
0 & 1 & 0 & 0\\
0 & 0 & 0 & 1\\
1 & 0 & 0 & 0
\end{array}\right), \quad 
K_0=\left(\begin{array}{cccc}
1 & 0 & 0 & 0\\
0 & 0 & 0 & 1\\
0 & 1 & 0 & 0\\
0 & 0 & 1 & 0
\end{array}\right),
\end{align}

\begin{figure}
\includegraphics*[width=0.45\textwidth]{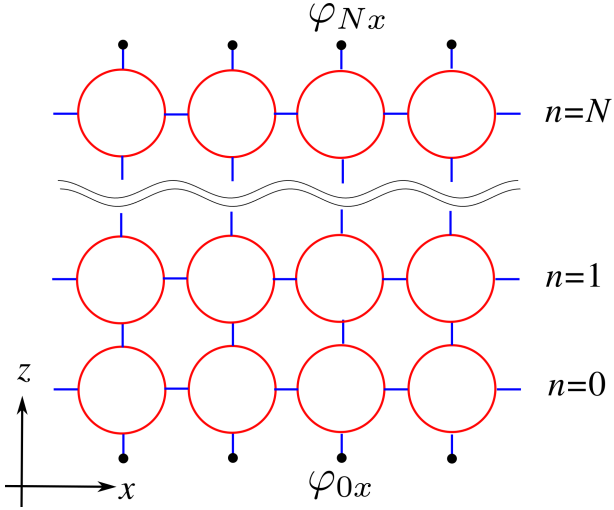} 
\caption{\label{Fig_boundary} (Color online) Cross sectional view of the slab 3d network model. It  has finite thickness in the $z$ direction, and has infinite extent in the $x$ and $y$ directions. The boundary rings at $n=0$ and $N$ has free legs. We introduce the phase delays $\varphi_{0x}$ and $\varphi_{Nx}$ there. 
 }  
\end{figure}

We define the slab surfaces in such a way that the $z$-oriented rings (rings perpendicular to the $z$ axis) are removed there. 
Then, the $x$- and $y$-oriented boundary rings have the free legs as shown in Fig. \ref{Fig_boundary}.  By introducing a phase delay there, we can control the boundary condition. The boundary condition thus becomes  
\begin{align}    
a_{0;x}'&=a_{0;x}e^{i\varphi_{0x}},& a_{0;y}&=c_{0;y}'e^{2i\phi}e^{i\varphi_{0y}},\label{Eq_boundary1}\\
b_{0;x}'&=b_{0;x}e^{i\varphi_{0x}},& d_{0;y}&=b_{0;y}'e^{2i\phi}e^{i\varphi_{0y}},\label{Eq_boundary2}\\
c_{N;x}'&=c_{N;x}e^{i\varphi_{Nx}},& b_{N;y}&=d_{N;y}'e^{2i\phi}e^{i\varphi_{Ny}},\label{Eq_boundary3}\\
d_{N;x}'&=d_{N;x}e^{i\varphi_{Nx}},& c_{N;y}&=a_{N;y}'e^{2i\phi}e^{i\varphi_{Ny}},\label{Eq_boundary4}
\end{align}
as shown in Fig. \ref{Fig_boundary}. 
Here, the phase delays are denoted as $\varphi_{0x},\varphi_{0y},\varphi_{Nx},\varphi_{Ny}$. They roughly corresponds to an elongation of the boundary rings.

\section{Symmetry}

As can be seen in Fig. \ref{Fig_Smatrix}, the S-matrix has various symmetries. 
If any non-reciprocal element is absent in the network, the time-reversal symmetry holds, resulting in 
\begin{align}
S_j^\dagger = \gamma_0 S_j^* \gamma_0, \quad 
\gamma_0=\left(\begin{array}{cccccccc}
0 & 1 & 0 & 0 & 0 & 0 & 0 & 0\\
1 & 0 & 0 & 0 & 0 & 0 & 0 & 0\\
0 & 0 & 0 & 1 & 0 & 0 & 0 & 0\\
0 & 0 & 1 & 0 & 0 & 0 & 0 & 0\\
0 & 0 & 0 & 0 & 0 & 1 & 0 & 0\\
0 & 0 & 0 & 0 & 1 & 0 & 0 & 0\\
0 & 0 & 0 & 0 & 0 & 0 & 0 & 1\\
0 & 0 & 0 & 0 & 0 & 0 & 1 & 0\\
\end{array}\right)
\end{align}
by assuming the unitarity. The parity symmetries with respect to $x$, $y$, and $z$ coordinates also hold. They give  
\begin{align}
&\gamma_0 S_x\gamma_0 = S_x,\quad \gamma_2S_y\gamma_2=S_y,\quad \gamma_1 S_z\gamma_1 = S_z,\\
&\gamma_1 S_x\gamma_1 = S_x,\quad \gamma_0S_y\gamma_0=S_y,\quad \gamma_2S_z \gamma_2=S_z,\\
&\gamma_2 S_x\gamma_2 = S_x,\quad \gamma_1S_y\gamma_1=S_y,\quad \gamma_0S_z\gamma_0=S_z,\\
&\gamma_1=\left(\begin{array}{cccccccc}
0 & 0 & 0 & 1 & 0 & 0 & 0 & 0\\
0 & 0 & 1 & 0 & 0 & 0 & 0 & 0\\
0 & 1 & 0 & 0 & 0 & 0 & 0 & 0\\
1 & 0 & 0 & 0 & 0 & 0 & 0 & 0\\
0 & 0 & 0 & 0 & 1 & 0 & 0 & 0\\
0 & 0 & 0 & 0 & 0 & 1 & 0 & 0\\
0 & 0 & 0 & 0 & 0 & 0 & 1 & 0\\
0 & 0 & 0 & 0 & 0 & 0 & 0 & 1\\
\end{array}\right),\\
&\gamma_2=\left(\begin{array}{cccccccc}
1 & 0 & 0 & 0 & 0 & 0 & 0 & 0\\
0 & 1 & 0 & 0 & 0 & 0 & 0 & 0\\
0 & 0 & 1 & 0 & 0 & 0 & 0 & 0\\
0 & 0 & 0 & 1 & 0 & 0 & 0 & 0\\
0 & 0 & 0 & 0 & 0 & 0 & 1 & 0\\
0 & 0 & 0 & 0 & 0 & 0 & 0 & 1\\
0 & 0 & 0 & 0 & 1 & 0 & 0 & 0\\
0 & 0 & 0 & 0 & 0 & 1 & 0 & 0\\
\end{array}\right). 
\end{align}
The $90^\circ$ rotational symmetry with respect to the $x$ axis results in 
\begin{align}
&\gamma_3^{-1}S_x\gamma_3 = S_x, \quad \gamma_3^{-1} S_y\gamma_3=S_z,\\
&\gamma_3=\left(\begin{array}{cccccccc}
0 & 0 & 0 & 0 & 0 & 0 & 0 & 1\\
0 & 0 & 0 & 0 & 0 & 0 & 1 & 0\\
0 & 0 & 0 & 0 & 1 & 0 & 0 & 0\\
0 & 0 & 0 & 0 & 0 & 1 & 0 & 0\\
0 & 1 & 0 & 0 & 0 & 0 & 0 & 0\\
1 & 0 & 0 & 0 & 0 & 0 & 0 & 0\\
0 & 0 & 1 & 0 & 0 & 0 & 0 & 0\\
0 & 0 & 0 & 1 & 0 & 0 & 0 & 0\\
\end{array}\right). 
\end{align}
Similarly, the rotational symmetry with respect to the $y$ and $z$ axes gives 
\begin{align}
&\gamma_3^{-1}S_y\gamma_3 = S_y, \quad \gamma_3^{-1} S_z\gamma_3=S_x,\\
&\gamma_3^{-1}S_z\gamma_3 = S_z, \quad \gamma_3^{-1} S_x\gamma_3=S_y,
\end{align}

Under these constraints by the symmetries, the S-matrices $S_x,S_y$ and $S_z$ must be the same, and are written as $S_j=\exp(iH)$ with 
\begin{align}
H=\left(\begin{array}{cccccccc}
a & b & c & d & e & f & e & f\\
b & a & d & c & f & e & f & e\\
c & d & a & b & f & e & f & e\\
d & c & b & a & e & f & e & f\\
e & f & f & e & a & b & d & c\\
f & e & e & f & b & a & c & d\\
e & f & f & e & d & c & a & b\\
f & e & e & f & c & d & b & a
\end{array}\right) \label{Eq_Hfull}
\end{align}
where $a,b,c,d,e,f$ are real parameter. The S-matrix itself has a similar expression as in Eq. (\ref{Eq_Hfull}), though the parameters are now complex.

Among the parameters, parameter $a$ is absorbed into redefinition of quasienergy $\phi$ for the bulk eigenvalue equation. 
However, it affects the slab modes through the boundary condition. Physical meanings of the other parameters are rather involved. However, if one of these parameters is nonzero, its meaning is clear.  Parameter $b$ represents the back scattering inside a ring.  Parameter $c$ describes the in-plane scattering of the same propagation direction. Namely, a clockwise mode in a ring is scattered into the clockwise modes in the nearest neighbor rings on the same plane.  Parameter $d$ describes the in-plane scattering of the opposite propagation direction. 
A clockwise mode is scattered into counter-clockwise modes. Parameters $e$ and $f$ describe the out-of-plane scattering. The former is reflection-like. If we inject via $a_{{\bm r};z}$ in Fig. \ref{Fig_Smatrix}, nonzero $a_{{\bm r};y}e^{-i\phi}$ and $b_{{\bm r}-\hat{z};y}e^{-i\phi}$ are  obtained with $e$. The latter is transmission-like. Nonzero $d_{{\bm r};y}e^{-i\phi}$ and $c_{{\bm r}-\hat{z};y}e^{-i\phi}$ are obtained with $f$ for the  $a_{{\bm r};z}$ injection.

These symmetries of the S-matrices yield the symmetry of the eigenvalue equation.
In terms of the equation, the time-reversal symmetry ${\cal T}$ is expressed as 
\begin{widetext}
\begin{align}
&{\cal T}U_{\bm k}^{-1}{\cal T}^{-1}=U_{-{\bm k}},\quad \phi(-{\bm k})=\phi({\bm k}),\\  
&{\cal T}={\cal V}^{-1}{\cal Q}{\cal G}_0{\cal V}{\cal K},\\
&{\cal Q}=\left(\begin{array}{cccccc}
(S_y^{-1})^{++} & 0 & 0 & 0 & 0 & (S_y^{-1})^{+-}\\
0 & (S_z^{-1})^{++} & 0 & (S_z^{-1})^{+-} & 0 & 0\\
0 & 0 & (S_x^{-1})^{++} & 0 & (S_x^{-1})^{+-} & 0\\
0 & (S_z^{-1})^{-+} & 0 & (S_z^{-1})^{--} & 0 & 0\\
0 & 0 & (S_x^{-1})^{-+} & 0 & (S_x^{-1})^{--} & 0\\
(S_y^{-1})^{-+} & 0 & 0 & 0 & 0 & (S_y^{-1})^{--}
\end{array}\right),\\
&{\cal V}={\rm Bdiag}(P_z^*,P_x^*,P_y^*,M_y^*,M_z^*,M_x^*),\\
&{\cal G}_0={\rm Bdiag}(G_0,G_0,G_0,G_0,G_0,G_0),\\
&G_0=\left(\begin{array}{cccc}
0 & 1 & 0 & 0\\
1 & 0 & 0 & 0\\
0 & 0 & 0 & 1\\
0 & 0 & 1 & 0
\end{array}\right),
\end{align}
\end{widetext}
where ${\cal K}$ is the complex conjugation operator and ``Bdiag'' stands for block diagonal.  
The time-reversal symmetry is bosonic, satisfying ${\cal T}^2=1$. 
Because of the bosonic property, the system will not be a 3d photonic TI with helical surface states.  There, the Kramers degeneracy due to the fermionic time-reversal symmetry plays a crucial role. 
With the same reason, the 2d network model with spin flipping cannot be a 2d photonic TI with helical edge states.   
Instead, the 3d network model may become a 3d photonic TCI whose robustness is protected by a spatial symmetry of the crystal structure. Therefore, it is important to investigate the spatial symmetry of the system.

The parity symmetries with respect to $x$ and $y$ coordinates are expressed as 
\begin{widetext}
\begin{align}
&{\cal P}_x U_{k_x,k_y}{\cal P}_x^{-1}=U_{-k_x,k_y}, \quad 
{\cal P}_y U_{k_x,k_y}{\cal P}_y^{-1}=U_{k_x,-k_y}, \quad 
\phi(-k_x,k_y)=\phi(k_x,-k_y)=\phi({\bm k}), \\
&{\cal P}_x= {\rm Bdiag}(1,G_1e^{-ik_x},G_0e^{-ik_x},1,G_1e^{-ik_x},G_0e^{-ik_x}),\\
&{\cal P}_y= {\rm Bdiag}(G_0e^{-ik_y},1,G_1e^{-ik_y},G_0e^{-ik_y},1,G_1e^{-ik_y}),\\
&G_1=\left(\begin{array}{cccc}
0 & 0 & 0 & 1\\
0 & 0 & 1 & 0\\
0 & 1 & 0 & 0\\
1 & 0 & 0 & 0
\end{array}\right).
\end{align}
The $90^\circ$ rotational symmetry with respect the $z$ axis is written as 
\begin{align}
&{\cal C}_4U_{\bm k}{\cal C}_4^{-1}=U_{-k_y,k_x}, \quad \phi(-k_y,k_x)=\phi({\bm k}),\\
&{\cal C}_4=\left(\begin{array}{cccccc}
0 & 0 & 0 & 0 & G_2 & 0\\
0 & 0 & 0 & G_3e^{-ik_y} & 0 & 0\\
0 & 0 & 0 & 0 & 0 & G_4e^{-ik_y}\\
0 & G_2^{-1} & 0 & 0 & 0 & 0\\
G_4e^{-ik_y} & 0 & 0 & 0 & 0 & 0\\
0 & 0 & G_3e^{-ik_y} & 0 & 0 & 0
\end{array}\right),\\
&G_2=\left(\begin{array}{cccc}
0 & 1 & 0 & 0\\
0 & 0 & 1 & 0\\
0 & 0 & 0 & 1\\
1 & 0 & 0 & 0
\end{array}\right), \quad 
G_3=\left(\begin{array}{cccc}
1 & 0 & 0 & 0\\
0 & 0 & 0 & 1\\
0 & 0 & 1 & 0\\
0 & 1 & 0 & 0
\end{array}\right), \quad 
G_4=\left(\begin{array}{cccc}
0 & 0 & 1 & 0\\
0 & 1 & 0 & 0\\
1 & 0 & 0 & 0\\
0 & 0 & 0 & 1
\end{array}\right). 
\end{align}
\end{widetext}
Obviously, ${\cal P}_x{\cal P}_y={\cal C}_4^2$ is satisfied. 
Other symmetry operations such as three-fold rotations, e.g., $(x,y,z)$ to $(y,z,x)$, can be constructed similarly. 
The system has totally the $O_h^1$ symmetry of the simple cubic lattice.

Besides, the system has the ``shift'' symmetry, if the off-plane scattering in $S_x$ is absent. For instance, if $S_x^{+-}=S_x^{-+}=0$, we have 
\begin{align}
&{\cal S}U_{\bm k}{\cal S}^{-1}=-U_{\bm k},\\
&{\cal S}=\pm{\rm Bdiag}(1,-1,-1,-1,1,1),
\end{align}
The shift symmetry implies that the quasienergy eigenvalues emerge as pairs of $\phi$ and $\phi+\pi$. Therefore, the quasienergy spectrum is invariant under $\pi$ shift. 
A similar invariance is obtained if $S_y^{+-}=S_y^{-+}=0$  or if $S_z^{+-}=S_z^{-+}=0$.  
This symmetries are irrespective of the spatial symmetry. 
The shift symmetry is enlarged if the off-plane  scattering is absent in all the directions. In this case, the relevant operator can be either one of the following forms. 
\begin{align}
{\cal S}=&\pm{\rm Bdiag}(1,1,1,-1,-1,-1),\\
&\pm{\rm Bdiag}(1,-1,1,-1,1,-1),  \\
&\pm{\rm Bdiag}(1,1,-1,-1,-1,1), \\
&\pm{\rm Bdiag}(1,-1,-1,-1,1,1.  
\end{align}
Although the vanishing off-plane scattering is unusual, we may have small off-plane scattering in an optical realization of the 3d network model. Therefore, it is reasonable to consider the  above situation as an extreme case.  

The symmetry properties of the eigenvalue equation constrain possible 
eigenmodes in bulk and slab systems, numerically obtained in the next section.

\section{Numerical results and discussions}

\begin{figure*}
\includegraphics*[width=0.95\textwidth]{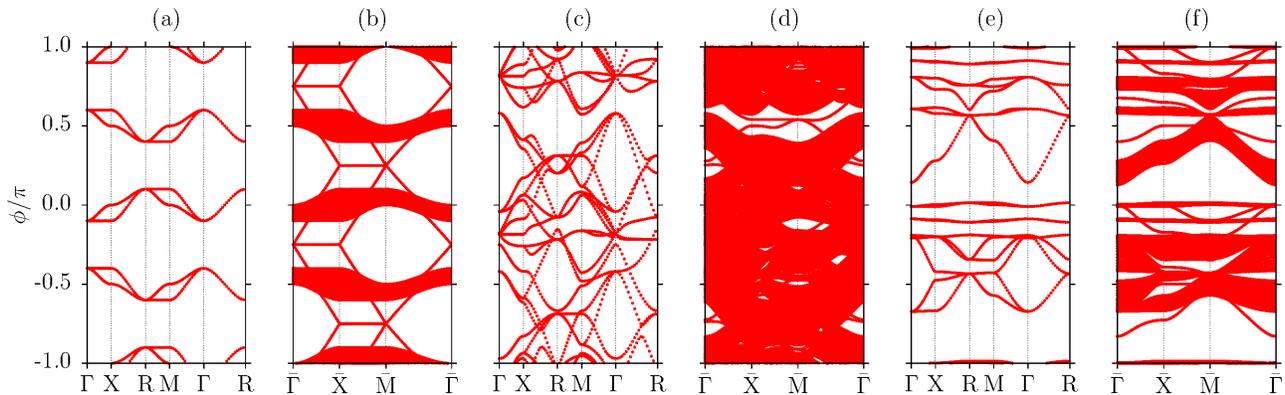} 
\caption{\label{Fig_band} (Color online) Bulk and slab band structures of the quasienergy in the 3d network model. Number of slab layers ($N$) is 16. Additional phases at the boundary rings ($\varphi_{0x},\varphi_{0y},\varphi_{Nx},\varphi_{Ny}$) are zero.     
The following S-matrix parameters are employed; $a=b=d=e=f=0$ and $c=0.4\pi$ [(a) and (b)].  $a=0.6117$, $b=0.5876$, $c=0.9623$, $d=0.8934$, $e=0.2022$, and $f=0.01273$, all multiplied by $2\pi$ [(c) and (d)], $a=0.9778$, $b=0.9019$, $c=0.6579$, $d=0.7289$, $e=0.4025$, and $f=0.9286$, all multiplied by $2\pi$ [(e) and (f)]. The latter two sets are generated randomly within (0,1).  
}
\end{figure*}

Let us numerically examine the quasienergy spectrum for the bulk and slab systems. 
In what follows, we assume the $O_h^1$ symmetry, so that $S_x=S_y=S_z$. 
Since there are the six free parameters in the S-matrix as shown in Eq. (\ref{Eq_Hfull}), we have a variety of the band structures depending on the parameter set. The bulk and slab band structures for three sets of the parameters are shown in Fig. \ref{Fig_band}.

We have 24 bulk eigenstates for a given ${\bm k}$, because the size of matrix $U_{\bm k}$ is $24\times 24$. The eigenvalues of $\phi$ as a function of ${\bm k}$ forms the band structure of 24 curves in the interval of $(-\pi,\pi)$. 
As for the slab system, we have $24N+16$ eigenmodes for a given ${\bm k}_\|$, momentum parallel to the slab surface. Here,  $N$ is the number of the ring layers.  In the  slab system, the eigenmodes  in the bulk band gaps are the surface states.  The surface  states are almost doubly degenerate, because the slab system has the two surfaces (the top and bottom ones), and the coupling between them are negligible. The surface states are evanescent away from the surface.

In Figs. \ref{Fig_band} (a) and (b), we assume an extreme case: Solely parameter $c$ in Eq. (\ref{Eq_Hfull}) is nonzero, and the rest parameters are zero. In this case, the system does not have the out-of-plane scattering and is merely the copies of the 2d spin-decoupled network \cite{PhysRevLett.110.203904}.  
As a result, the shift symmetry becomes exact, so that the bulk band structure is invariant under the $\pi$ shift, $\phi\to\phi+\pi$. 
Let us remind that parameter $c$ specifies the in-plane coupling to the adjacent ring with the same propagation direction. 
According to the result of the 2d network \cite{PhysRevLett.110.203904}, if $c>\pi/4$, the system is in the anomalous Floquet-insulator phase, and exhibits gapless edge states.  
The parameter used in the numerical calculation satisfies this inequality, so that gapless surface states are found.  
We can see the completely flat dispersion of the surface bands along the $\bar{\Gamma}\bar{\rm X}$ and $\bar{\rm X}\bar{\rm M}$ directions. This is because the off-plane interaction is absent with the parameter set.

In Figs. \ref{Fig_band} (c) and (d), all the parameters in Eq. (\ref{Eq_Hfull}) are nonzero, and were generated randomly within $(0,2\pi)$.  The band structure  exhibits the gapless surface states in the bulk band gap around $\phi/\pi=0.5$.
The surface-state dispersion curves touch quadratically at the $\bar{\textrm{M}}$ point in the surface Brillouin zone. This property is reminiscent of the surface states in TCI \cite{fu2011topological}.

In Figs. \ref{Fig_band} (e) and (f), all the parameters are again generated randomly, and the band structures are completely different from those in \ref{Fig_band} (c) and (d).  In this way, depending on the parameter set, a wide variety of gapless and gapped band structures both in the bulk and slab systems are available. In this case, we can observe nearly-flat bulk bands around $\phi/\pi=-0.1$, 0.0, 0.9, and 1.0. Note that the flat bands satisfy the shift symmetry approximately, though the other bands do not. 
In the vicinity of the flat bands, the gapless surface states are observed.
The other surface states found around $\phi/\pi=-0.8$ and 0.5 are gapped and merge with bulk bands.

\begin{figure*}
\includegraphics*[width=0.95\textwidth]{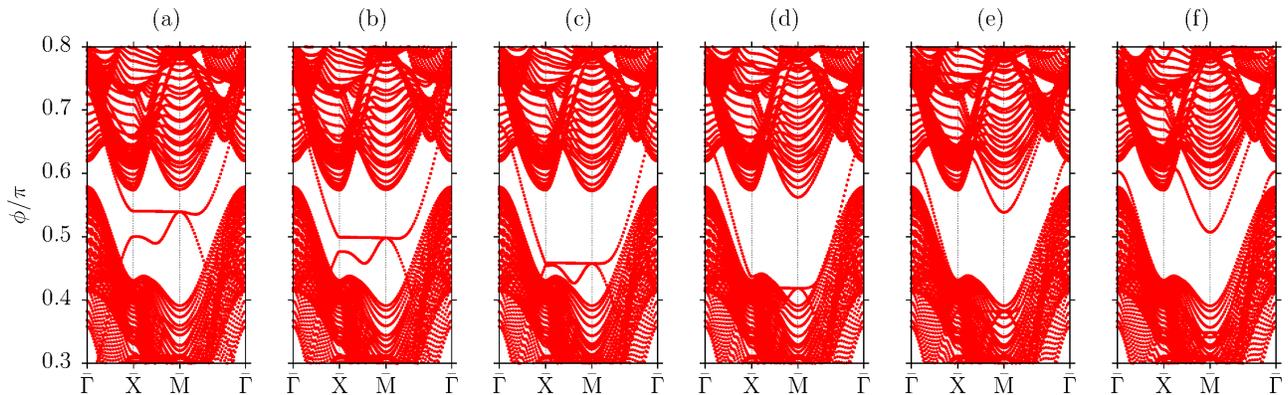}  
\caption{\label{Fig_robust} (Color online) Surface band structures versus the phase delay of the surface rings.  Parameters are the same as in Fig. \ref{Fig_band} (d), except for the phase delay $\varphi_d\equiv\varphi_{0x}=\varphi_{0y}=\varphi_{Nx}=\varphi_{Ny}$. The phase delay is 0 (a), $0.1\pi$ (b), $0.2\pi$ (c), $0.3\pi$ (d), $0.4\pi$ (e), and $0.5\pi$ (f).
}
\end{figure*}

An important question here is whether these gapless surface states are topological or not. In other words, are they robust against certain perturbations? 
To answer this question, we modify the boundary condition by introducing nonzero phase delays $\varphi_{0x}=\varphi_{0y}=\varphi_{Nx}=\varphi_{Ny}\equiv\varphi_d$.  The result is shown in  Fig. \ref{Fig_robust} for the gapless surface states in Fig. \ref{Fig_band} (d).
We can see that the gapless surface states hold for $\varphi_d<0.4\pi$, preserving the quadratic degeneracy at $\bar{\textrm{M}}$. However, they are gapped out at 
$\varphi_d=0.5\pi$. Therefore, the surface states are not robust against the large phase delay. 
For comparison, the gapless surface states in Fig. \ref{Fig_band} (b) are robust against the phase delay. The gaplessness holds for any  $\varphi_d$. 
Those in  Fig. \ref{Fig_band} (f) are fragile. Even at $\varphi_d=0.1\pi$, the upper surface states around $\phi/\pi=0.9$ are gapped out. At $\varphi_d=0.3\pi$, the lower surface states around  $\phi/\pi=0.7$ are gapped out too (not shown).

As mentioned, the dispersion curve of the gapless surface states in Fig. \ref{Fig_band} (d) is similar to that of the TCI of Ref. \onlinecite{fu2007topological}. However, the robustness is partial in our case. 
In our system, we have the $O_h^1$ and $C_{4v}$ symmetries in the bulk and surface, respectively. Therefore, the bosonic time-reversal symmetry (${\cal T}$) along with $90^\circ$ rotation ($C_4$) derived in Sec. III are shared with the TCI in common. However, the $O_h^1$ symmetry and its resulting triply-degenerate modes at the $\Gamma$ and R points in the Brillouin zone 
prohibits the $Z_2$ characterization of the TCI. Thus, the $Z_2$ topology is invisible.

Besides, our system is also characterized as  a Floquet system. In such a system, a topological invariant is often given by  
the homotopy invariants as \cite{PhysRevB.82.235114}
\begin{align}
&\nu_i^{(1)}=\int \frac{dk_i}{2\pi}\textrm{tr}(U_{\bm k}^\dagger\partial_iU_{\bm k}),\\
&\nu^{(3)}=\int \frac{d^3k}{24\pi^2}\sum_{ijk}\epsilon_{ijk}\textrm{tr}(U_{\bm k}^\dagger\partial_iU_{\bm k} U_{\bm k}^\dagger\partial_jU_{\bm k} U_{\bm k}^\dagger\partial_kU_{\bm k} ).
\end{align}
The former maps from $T^1$ (1d section of the 3d Brillouin zone) to  $U(24)$ (unitary matrix $U_{\bm k}$). The latter maps from 
$T^3$ (the 3d Brillouin zone) to $U(24)$. 
However, by explicit calculation, they are shown to be vanishing in our system.

We are thus forced to seek other topological invariants that classify possible topological phases in our system.  
Our system is attributed in the AI class in the Altland-Zirnbauer symmetry classes \cite{PhysRevB.55.1142}.  Namely, the  bosonic time-reversal symmetry exists, but the particle-hole and chiral symmetries are absent.  According to the periodic table of TI \cite{schnyder2008classification}, there is no topological invariant in 3d systems, protected solely by the on-site  symmetries (the time-reversal, charge-conjugation, and chiral symmetries).  
Recently, the periodic table is extended to deal with topological phases protected by the on-site symmetries and spatial symmetries \cite{PhysRevB.88.125129,PhysRevB.90.165114}. 
However, to the best of our knowledge, topological phases inherent in Floquet systems  protected by these symmetries too are still uncovered in the existing periodic tables.

For comparison, the 2d spin-decoupled network model is attributed to the AIII class for each spin sector. The periodic table indicates that the topology is trivial. However, it is not the case.

\section{Summary and Outlook}

In summary, we have presented fundamental properties of the 3d Chalker-Coddington type network model with the spatial periodicity of the simple cubic lattice. 
It maps a Floquet-Bloch system, and the quasienergy spectrum exhibits a 3d band structure.   
Depending on the S-matrices among the ring-propagation modes, the system can have gapped bulk band structures and gapless surface states. A close resemblance to the TCI is drawn. However, the topological characterization of the model is yet unclarified.

\begin{figure}
\includegraphics*[width=0.45\textwidth]{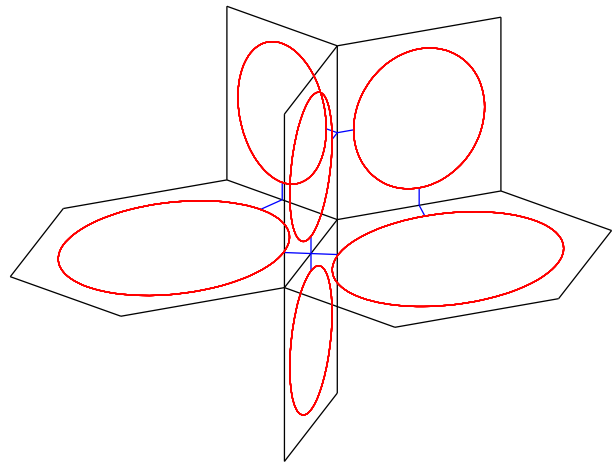} 
\caption{\label{Fig_hexagonal} (Color online) Schematic illustration of 3d hexagonal network model.}
\end{figure}

Although we consider the simple cubic lattice with the $O_h^1$ symmetry, an interesting opportunity may arise when we consider other types of 3d lattice structures. In the 2d networks, the hexagonal geometry gives rise to both the Chern-insulator  and anomalous Floquet-insulators phases, whereas in the square-lattice network, the Chern-insulator phase is absent \cite{PhysRevB.89.075113}. Thus, the 3d hexagonal lattice is an interesting option. 
In this case, relevant S-matrices are forced to be anisotropic. 
They have the sizes $8\times8$ and $6\times 6$ for planar and vertical hinges, respectively, as shown in Fig. \ref{Fig_hexagonal}
The spatial symmetry of the 3d hexagonal network model is $D_{6h}$. 
Compared with $O_h^1$ of the simple cubic case, the $D_{6h}$ point group does not support triply degenerate modes without accidental conditions.

Changing system symmetries are also available by elongating $x$ and $y$-oriented rings in the simple cubic lattice, or by introducing nonreciprocal elements in the network.  In the former case, the relevant symmetry is $D_{4h}$, while in the latter, the time-reversal symmetry no longer holds. Since the spatial and time-reversal symmetries are critical in such systems, a variety of novel phenomena  are expected to occur. However, we should keep in mind that reduced symmetries 
tend to close possible band gaps in the bulk, because possibly degeneracies in the quasienergy are lifted. This may hide the gapless surface states.

\begin{figure}
\includegraphics*[width=0.45\textwidth]{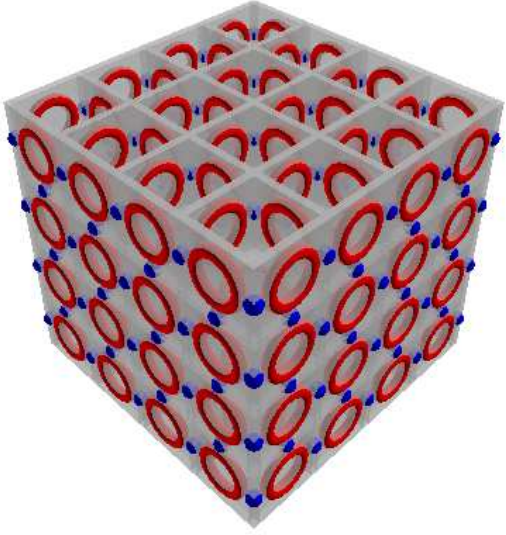} 
\caption{\label{Fig_realization} (Color online) Optical realization of the 3d network model. It consists of high-index ring-resonator arrays placed on a low-index circuit board. At the hinges of the board lattice, high-index spherical resonators are embedded.  
}
\end{figure}

We should comment on how to realize the 3d network model.  
A possible optical realization  is a 3d ring-resonator lattice accompanied by spherical resonator placed among the adjacent rings, as shown in Fig. \ref{Fig_realization}. 
As in the 3d metamaterial of  split-ring resonators and wires \cite{PhysRevLett.84.4184}, the simple cubic lattice composed of low-index circuit boards are prepared. The square-lattice of high-index ring network is printed on the boards. High-index  spherical resonators are embedded on  every hinge of the board lattice. The spherical resonators in the x-oriented hinges  can be replaced by x-oriented ring resonators.  
The S-matrix parameters can be controlled by the geometry and resonator profiles. For instance, at a frequency near a resonance of the hinge resonators, the coupling among the ring resonators is strongly enhanced. This will result in gapped bulk band structures and gapless surface states as given in Sec. IV. Actual design of the ring and hinge resonators are beyond the scope of the present paper.

With the construction, a synthetic gauge field for photons may be implemented by tailoring the hinge resonators. We can show that the network model is invariant under the following gauge transformation:
\begin{align}
& a_{{\bm r};x}\to a_{{\bm r};x}e^{i\theta({\bm r})},\\
& b_{{\bm r};x}\to b_{{\bm r};x}e^{i\theta({\bm r}+\hat{y})},\\
& c_{{\bm r};x}\to c_{{\bm r};x}e^{i\theta({\bm r}+\hat{y}+\hat{z})},\\
& d_{{\bm r};x}\to d_{{\bm r};x}e^{i\theta({\bm r}+\hat{z})},\\
& a'_{{\bm r};x}\to a'_{{\bm r};x}e^{i\theta({\bm r}+\hat{y})},\\
& b'_{{\bm r};x}\to b'_{{\bm r};x}e^{i\theta({\bm r})},\\
& c'_{{\bm r};x}\to c'_{{\bm r};x}e^{i\theta({\bm r}+\hat{z})},\\
& d'_{{\bm r};x}\to d'_{{\bm r};x}e^{i\theta({\bm r}+\hat{y}+\hat{z})},\\
\end{align}
accompanied by the cyclic rotation of Cartesian indices for the other components of the mode amplitudes.   
The gauge field is introduced in the  S-matrices as 
\begin{widetext}
\begin{align}
&S_{{\bm r}\mu}(A)= \left(\begin{array}{llllllll}
S_\mu^{11} e^{iA_{{\bm r}\mu}}& S_\mu^{12} & S_\mu^{13} & S_\mu^{14} e^{iA_{{\bm r}\mu}} & S_\mu^{15} & S_\mu^{16} e^{iA_{{\bm r}\mu}} & S_\mu^{17} & S_\mu^{18} e^{iA_{{\bm r}\mu}}\\
S_\mu^{21} & S_\mu^{22}e^{-iA_{{\bm r}\mu}}& S_\mu^{23} e^{-iA_{{\bm r}\mu}} & S_\mu^{24} & S_\mu^{25} e^{-iA_{{\bm r}\mu}} & S_\mu^{26} & S_\mu^{27} e^{-iA_{{\bm r}\mu}} & S_\mu^{28}\\
S_\mu^{31} & S_\mu^{32}e^{-iA_{{\bm r}\mu}}& S_\mu^{33} e^{-iA_{{\bm r}\mu}} & S_\mu^{34} & S_\mu^{35} e^{-iA_{{\bm r}\mu}} & S_\mu^{36} & S_\mu^{37} e^{-iA_{{\bm r}\mu}} & S_\mu^{38}\\
S_\mu^{41} e^{iA_{{\bm r}\mu}}& S_\mu^{42} & S_\mu^{43} & S_\mu^{44} e^{iA_{{\bm r}\mu}} & S_\mu^{45} & S_\mu^{46} e^{iA_{{\bm r}\mu}} & S_\mu^{47} & S_\mu^{48} e^{iA_{{\bm r}\mu}}\\
S_\mu^{51} & S_\mu^{52}e^{-iA_{{\bm r}\mu}}& S_\mu^{53} e^{-iA_{{\bm r}\mu}} & S_\mu^{54} & S_\mu^{55} e^{-iA_{{\bm r}\mu}} & S_\mu^{56} & S_\mu^{57} e^{iA_{{\bm r}\mu}} & S_\mu^{58}\\
S_\mu^{61} e^{iA_{{\bm r}\mu}}& S_\mu^{62} & S_\mu^{63} & S_\mu^{64} e^{iA_{{\bm r}\mu}} & S_\mu^{65} & S_\mu^{66} e^{iA_{{\bm r}\mu}} & S_\mu^{67} & S_\mu^{68} e^{iA_{{\bm r}\mu}}\\
S_\mu^{71} & S_\mu^{72}e^{-iA_{{\bm r}\mu}}& S_\mu^{73} e^{-iA_{{\bm r}\mu}} & S_\mu^{74} & S_\mu^{75} e^{-iA_{{\bm r}\mu}} & S_\mu^{76} & S_\mu^{77} e^{-iA_{{\bm r}\mu}} & S_\mu^{78}\\
S_\mu^{81} e^{iA_{{\bm r}\mu}}& S_\mu^{82} & S_\mu^{83} & S_\mu^{84} e^{iA_{{\bm r}\mu}} & S_\mu^{85} & S_\mu^{86} e^{iA_{{\bm r}\mu}} & S_\mu^{87} & S_\mu^{88} e^{iA_{{\bm r}\mu}}
\end{array}\right),
\end{align}
\end{widetext}
where the gauge transformation is defined as 
\begin{align}
A_{{\bm r};\mu} \to A_{{\bm r};\mu}+\theta({\bm r}+\hat{\mu})-\theta({\bm r}). 
\end{align}
The gauge field (gauge connection) on the lattice is naturally placed on the hinge. 
By introducing different gauge fields on different hinges, we can investigate various phenomena of photons under synthetic gauge fields. The three-dimensional Hofstadter butterfly \cite{PhysRevB.45.13488,PhysRevLett.86.1062} may be realized optically in the network model.

We hope this paper stimulates further investigation of the 3d Chalker-Coddington network model both in periodic and random setups, from various viewpoints such as photonic TIs, Anderson localization, synthetic gauge field, and cavity QED lattices.

\begin{acknowledgments}
This work was partially supported by JSPS KAKENHI (Grant No. 26390013). 
\end{acknowledgments}


\begin{thebibliography}{31}
\expandafter\ifx\csname natexlab\endcsname\relax\def\natexlab#1{#1}\fi
\expandafter\ifx\csname bibnamefont\endcsname\relax
  \def\bibnamefont#1{#1}\fi
\expandafter\ifx\csname bibfnamefont\endcsname\relax
  \def\bibfnamefont#1{#1}\fi
\expandafter\ifx\csname citenamefont\endcsname\relax
  \def\citenamefont#1{#1}\fi
\expandafter\ifx\csname url\endcsname\relax
  \def\url#1{\texttt{#1}}\fi
\expandafter\ifx\csname urlprefix\endcsname\relax\def\urlprefix{URL }\fi
\providecommand{\bibinfo}[2]{#2}
\providecommand{\eprint}[2][]{\url{#2}}

\bibitem[{\citenamefont{Hafezi et~al.}(2013)\citenamefont{Hafezi, Mittal, Fan,
  Migdall, and Taylor}}]{hafezi2013imaging}
\bibinfo{author}{\bibfnamefont{M.}~\bibnamefont{Hafezi}},
  \bibinfo{author}{\bibfnamefont{S.}~\bibnamefont{Mittal}},
  \bibinfo{author}{\bibfnamefont{J.}~\bibnamefont{Fan}},
  \bibinfo{author}{\bibfnamefont{A.}~\bibnamefont{Migdall}}, \bibnamefont{and}
  \bibinfo{author}{\bibfnamefont{J.~M.} \bibnamefont{Taylor}},
  \bibinfo{journal}{Nature Photon.} \textbf{\bibinfo{volume}{7}},
  \bibinfo{pages}{1001} (\bibinfo{year}{2013}).

\bibitem[{\citenamefont{Kane and Mele}(2005)}]{kane2005qsh}
\bibinfo{author}{\bibfnamefont{C.~L.} \bibnamefont{Kane}} \bibnamefont{and}
  \bibinfo{author}{\bibfnamefont{E.~J.} \bibnamefont{Mele}},
  \bibinfo{journal}{Phys. Rev. Lett.} \textbf{\bibinfo{volume}{95}},
  \bibinfo{pages}{226801} (\bibinfo{year}{2005}).

\bibitem[{\citenamefont{Liang and Chong}(2013)}]{PhysRevLett.110.203904}
\bibinfo{author}{\bibfnamefont{G.~Q.} \bibnamefont{Liang}} \bibnamefont{and}
  \bibinfo{author}{\bibfnamefont{Y.~D.} \bibnamefont{Chong}},
  \bibinfo{journal}{Phys. Rev. Lett.} \textbf{\bibinfo{volume}{110}},
  \bibinfo{pages}{203904} (\bibinfo{year}{2013}).

\bibitem[{\citenamefont{Pasek and Chong}(2014)}]{PhysRevB.89.075113}
\bibinfo{author}{\bibfnamefont{M.}~\bibnamefont{Pasek}} \bibnamefont{and}
  \bibinfo{author}{\bibfnamefont{Y.~D.} \bibnamefont{Chong}},
  \bibinfo{journal}{Phys. Rev. B} \textbf{\bibinfo{volume}{89}},
  \bibinfo{pages}{075113} (\bibinfo{year}{2014}).

\bibitem[{\citenamefont{Semenoff}(1984)}]{semenoff1984cms}
\bibinfo{author}{\bibfnamefont{G.~W.} \bibnamefont{Semenoff}},
  \bibinfo{journal}{Phys. Rev. Lett.} \textbf{\bibinfo{volume}{53}},
  \bibinfo{pages}{2449} (\bibinfo{year}{1984}).

\bibitem[{\citenamefont{Haldane}(1988)}]{haldane1988mqh}
\bibinfo{author}{\bibfnamefont{F.~D.~M.} \bibnamefont{Haldane}},
  \bibinfo{journal}{Phys. Rev. Lett.} \textbf{\bibinfo{volume}{61}},
  \bibinfo{pages}{2015} (\bibinfo{year}{1988}).

\bibitem[{\citenamefont{Kitagawa et~al.}(2010)\citenamefont{Kitagawa, Berg,
  Rudner, and Demler}}]{PhysRevB.82.235114}
\bibinfo{author}{\bibfnamefont{T.}~\bibnamefont{Kitagawa}},
  \bibinfo{author}{\bibfnamefont{E.}~\bibnamefont{Berg}},
  \bibinfo{author}{\bibfnamefont{M.}~\bibnamefont{Rudner}}, \bibnamefont{and}
  \bibinfo{author}{\bibfnamefont{E.}~\bibnamefont{Demler}},
  \bibinfo{journal}{Phys. Rev. B} \textbf{\bibinfo{volume}{82}},
  \bibinfo{pages}{235114} (\bibinfo{year}{2010}).

\bibitem[{\citenamefont{Chalker and Coddington}(1988)}]{0022-3719-21-14-008}
\bibinfo{author}{\bibfnamefont{J.~T.} \bibnamefont{Chalker}} \bibnamefont{and}
  \bibinfo{author}{\bibfnamefont{P.~D.} \bibnamefont{Coddington}},
  \bibinfo{journal}{J. Phys. C: Solid State Physics}
  \textbf{\bibinfo{volume}{21}}, \bibinfo{pages}{2665} (\bibinfo{year}{1988}).

\bibitem[{\citenamefont{Janssen et~al.}(1999)\citenamefont{Janssen, Metzler,
  and Zirnbauer}}]{PhysRevB.59.15836}
\bibinfo{author}{\bibfnamefont{M.}~\bibnamefont{Janssen}},
  \bibinfo{author}{\bibfnamefont{M.}~\bibnamefont{Metzler}}, \bibnamefont{and}
  \bibinfo{author}{\bibfnamefont{M.~R.} \bibnamefont{Zirnbauer}},
  \bibinfo{journal}{Phys. Rev. B} \textbf{\bibinfo{volume}{59}},
  \bibinfo{pages}{15836} (\bibinfo{year}{1999}).

\bibitem[{\citenamefont{Ho and Chalker}(1996)}]{PhysRevB.54.8708}
\bibinfo{author}{\bibfnamefont{C.-M.} \bibnamefont{Ho}} \bibnamefont{and}
  \bibinfo{author}{\bibfnamefont{J.~T.} \bibnamefont{Chalker}},
  \bibinfo{journal}{Phys. Rev. B} \textbf{\bibinfo{volume}{54}},
  \bibinfo{pages}{8708} (\bibinfo{year}{1996}).

\bibitem[{\citenamefont{Khanikaev et~al.}(2013)\citenamefont{Khanikaev,
  Mousavi, Tse, Kargarian, MacDonald, and Shvets}}]{khanikaev2013photonic}
\bibinfo{author}{\bibfnamefont{A.~B.} \bibnamefont{Khanikaev}},
  \bibinfo{author}{\bibfnamefont{S.~H.} \bibnamefont{Mousavi}},
  \bibinfo{author}{\bibfnamefont{W.-K.} \bibnamefont{Tse}},
  \bibinfo{author}{\bibfnamefont{M.}~\bibnamefont{Kargarian}},
  \bibinfo{author}{\bibfnamefont{A.~H.} \bibnamefont{MacDonald}},
  \bibnamefont{and} \bibinfo{author}{\bibfnamefont{G.}~\bibnamefont{Shvets}},
  \bibinfo{journal}{Nature Mater.} \textbf{\bibinfo{volume}{12}},
  \bibinfo{pages}{233} (\bibinfo{year}{2013}).

\bibitem[{\citenamefont{Chen et~al.}(2014)\citenamefont{Chen, Jiang, Chen, Zhu,
  Zhou, Dong, and Chan}}]{chen2014experimental}
\bibinfo{author}{\bibfnamefont{W.-J.} \bibnamefont{Chen}},
  \bibinfo{author}{\bibfnamefont{S.-J.} \bibnamefont{Jiang}},
  \bibinfo{author}{\bibfnamefont{X.-D.} \bibnamefont{Chen}},
  \bibinfo{author}{\bibfnamefont{B.}~\bibnamefont{Zhu}},
  \bibinfo{author}{\bibfnamefont{L.}~\bibnamefont{Zhou}},
  \bibinfo{author}{\bibfnamefont{J.-W.} \bibnamefont{Dong}}, \bibnamefont{and}
  \bibinfo{author}{\bibfnamefont{C.~T.} \bibnamefont{Chan}},
  \bibinfo{journal}{Nature Comm.} \textbf{\bibinfo{volume}{5}}
  (\bibinfo{year}{2014}).

\bibitem[{\citenamefont{Ochiai}(2015)}]{ochiai2015time}
\bibinfo{author}{\bibfnamefont{T.}~\bibnamefont{Ochiai}}, \bibinfo{journal}{J.
  Phys. Soc. Jpn.} \textbf{\bibinfo{volume}{84}}, \bibinfo{pages}{054401}
  (\bibinfo{year}{2015}).

\bibitem[{\citenamefont{Rechtsman et~al.}(2013)\citenamefont{Rechtsman, Zeuner,
  Plotnik, Lumer, Podolsky, Dreisow, Nolte, Segev, and
  Szameit}}]{rechtsman2013photonic}
\bibinfo{author}{\bibfnamefont{M.~C.} \bibnamefont{Rechtsman}},
  \bibinfo{author}{\bibfnamefont{J.~M.} \bibnamefont{Zeuner}},
  \bibinfo{author}{\bibfnamefont{Y.}~\bibnamefont{Plotnik}},
  \bibinfo{author}{\bibfnamefont{Y.}~\bibnamefont{Lumer}},
  \bibinfo{author}{\bibfnamefont{D.}~\bibnamefont{Podolsky}},
  \bibinfo{author}{\bibfnamefont{F.}~\bibnamefont{Dreisow}},
  \bibinfo{author}{\bibfnamefont{S.}~\bibnamefont{Nolte}},
  \bibinfo{author}{\bibfnamefont{M.}~\bibnamefont{Segev}}, \bibnamefont{and}
  \bibinfo{author}{\bibfnamefont{A.}~\bibnamefont{Szameit}},
  \bibinfo{journal}{Nature} \textbf{\bibinfo{volume}{496}},
  \bibinfo{pages}{196} (\bibinfo{year}{2013}).

\bibitem[{\citenamefont{Wu and Hu}(2015)}]{PhysRevLett.114.223901}
\bibinfo{author}{\bibfnamefont{L.-H.} \bibnamefont{Wu}} \bibnamefont{and}
  \bibinfo{author}{\bibfnamefont{X.}~\bibnamefont{Hu}}, \bibinfo{journal}{Phys.
  Rev. Lett.} \textbf{\bibinfo{volume}{114}}, \bibinfo{pages}{223901}
  (\bibinfo{year}{2015}).

\bibitem[{\citenamefont{Yannopapas}(2011)}]{PhysRevB.84.195126}
\bibinfo{author}{\bibfnamefont{V.}~\bibnamefont{Yannopapas}},
  \bibinfo{journal}{Phys. Rev. B} \textbf{\bibinfo{volume}{84}},
  \bibinfo{pages}{195126} (\bibinfo{year}{2011}).

\bibitem[{\citenamefont{Lu et~al.}(2015)\citenamefont{Lu, Fang, Fu, Johnson,
  Joannopoulos, and Solja{\v{c}}i{\'c}}}]{lu2015three}
\bibinfo{author}{\bibfnamefont{L.}~\bibnamefont{Lu}},
  \bibinfo{author}{\bibfnamefont{C.}~\bibnamefont{Fang}},
  \bibinfo{author}{\bibfnamefont{L.}~\bibnamefont{Fu}},
  \bibinfo{author}{\bibfnamefont{S.~G.} \bibnamefont{Johnson}},
  \bibinfo{author}{\bibfnamefont{J.~D.} \bibnamefont{Joannopoulos}},
  \bibnamefont{and}
  \bibinfo{author}{\bibfnamefont{M.}~\bibnamefont{Solja{\v{c}}i{\'c}}},
  \bibinfo{journal}{arXiv:1507.00337}  (\bibinfo{year}{2015}).

\bibitem[{\citenamefont{Obuse et~al.}(2007)\citenamefont{Obuse, Furusaki, Ryu,
  and Mudry}}]{PhysRevB.76.075301}
\bibinfo{author}{\bibfnamefont{H.}~\bibnamefont{Obuse}},
  \bibinfo{author}{\bibfnamefont{A.}~\bibnamefont{Furusaki}},
  \bibinfo{author}{\bibfnamefont{S.}~\bibnamefont{Ryu}}, \bibnamefont{and}
  \bibinfo{author}{\bibfnamefont{C.}~\bibnamefont{Mudry}},
  \bibinfo{journal}{Phys. Rev. B} \textbf{\bibinfo{volume}{76}},
  \bibinfo{pages}{075301} (\bibinfo{year}{2007}).

\bibitem[{\citenamefont{Ryu et~al.}(2010)\citenamefont{Ryu, Mudry, Obuse, and
  Furusaki}}]{1367-2630-12-6-065005}
\bibinfo{author}{\bibfnamefont{S.}~\bibnamefont{Ryu}},
  \bibinfo{author}{\bibfnamefont{C.}~\bibnamefont{Mudry}},
  \bibinfo{author}{\bibfnamefont{H.}~\bibnamefont{Obuse}}, \bibnamefont{and}
  \bibinfo{author}{\bibfnamefont{A.}~\bibnamefont{Furusaki}},
  \bibinfo{journal}{New J. Phys.} \textbf{\bibinfo{volume}{12}},
  \bibinfo{pages}{065005} (\bibinfo{year}{2010}).

\bibitem[{\citenamefont{Fu}(2011)}]{fu2011topological}
\bibinfo{author}{\bibfnamefont{L.}~\bibnamefont{Fu}}, \bibinfo{journal}{Phys.
  Rev. Lett.} \textbf{\bibinfo{volume}{106}}, \bibinfo{pages}{106802}
  (\bibinfo{year}{2011}).

\bibitem[{\citenamefont{Chalker and Dohmen}(1995)}]{PhysRevLett.75.4496}
\bibinfo{author}{\bibfnamefont{J.~T.} \bibnamefont{Chalker}} \bibnamefont{and}
  \bibinfo{author}{\bibfnamefont{A.}~\bibnamefont{Dohmen}},
  \bibinfo{journal}{Phys. Rev. Lett.} \textbf{\bibinfo{volume}{75}},
  \bibinfo{pages}{4496} (\bibinfo{year}{1995}).

\bibitem[{\citenamefont{Klesse and Metzler}(1999)}]{klesse1999modeling}
\bibinfo{author}{\bibfnamefont{R.}~\bibnamefont{Klesse}} \bibnamefont{and}
  \bibinfo{author}{\bibfnamefont{M.}~\bibnamefont{Metzler}},
  \bibinfo{journal}{Int. J. Mod. Phys. C} \textbf{\bibinfo{volume}{10}},
  \bibinfo{pages}{577} (\bibinfo{year}{1999}).

\bibitem[{\citenamefont{Obuse et~al.}(2014)\citenamefont{Obuse, Ryu, Furusaki,
  and Mudry}}]{PhysRevB.89.155315}
\bibinfo{author}{\bibfnamefont{H.}~\bibnamefont{Obuse}},
  \bibinfo{author}{\bibfnamefont{S.}~\bibnamefont{Ryu}},
  \bibinfo{author}{\bibfnamefont{A.}~\bibnamefont{Furusaki}}, \bibnamefont{and}
  \bibinfo{author}{\bibfnamefont{C.}~\bibnamefont{Mudry}},
  \bibinfo{journal}{Phys. Rev. B} \textbf{\bibinfo{volume}{89}},
  \bibinfo{pages}{155315} (\bibinfo{year}{2014}).

\bibitem[{\citenamefont{Fu et~al.}(2007)\citenamefont{Fu, Kane, and
  Mele}}]{fu2007topological}
\bibinfo{author}{\bibfnamefont{L.}~\bibnamefont{Fu}},
  \bibinfo{author}{\bibfnamefont{C.~L.} \bibnamefont{Kane}}, \bibnamefont{and}
  \bibinfo{author}{\bibfnamefont{E.~J.} \bibnamefont{Mele}},
  \bibinfo{journal}{Phys. Rev. Lett.} \textbf{\bibinfo{volume}{98}},
  \bibinfo{pages}{106803} (\bibinfo{year}{2007}).

\bibitem[{\citenamefont{Altland and Zirnbauer}(1997)}]{PhysRevB.55.1142}
\bibinfo{author}{\bibfnamefont{A.}~\bibnamefont{Altland}} \bibnamefont{and}
  \bibinfo{author}{\bibfnamefont{M.~R.} \bibnamefont{Zirnbauer}},
  \bibinfo{journal}{Phys. Rev. B} \textbf{\bibinfo{volume}{55}},
  \bibinfo{pages}{1142} (\bibinfo{year}{1997}).

\bibitem[{\citenamefont{Schnyder et~al.}(2008)\citenamefont{Schnyder, Ryu,
  Furusaki, and Ludwig}}]{schnyder2008classification}
\bibinfo{author}{\bibfnamefont{A.~P.} \bibnamefont{Schnyder}},
  \bibinfo{author}{\bibfnamefont{S.}~\bibnamefont{Ryu}},
  \bibinfo{author}{\bibfnamefont{A.}~\bibnamefont{Furusaki}}, \bibnamefont{and}
  \bibinfo{author}{\bibfnamefont{A.~W.} \bibnamefont{Ludwig}},
  \bibinfo{journal}{Phys. Rev. B} \textbf{\bibinfo{volume}{78}},
  \bibinfo{pages}{195125} (\bibinfo{year}{2008}).

\bibitem[{\citenamefont{Morimoto and Furusaki}(2013)}]{PhysRevB.88.125129}
\bibinfo{author}{\bibfnamefont{T.}~\bibnamefont{Morimoto}} \bibnamefont{and}
  \bibinfo{author}{\bibfnamefont{A.}~\bibnamefont{Furusaki}},
  \bibinfo{journal}{Phys. Rev. B} \textbf{\bibinfo{volume}{88}},
  \bibinfo{pages}{125129} (\bibinfo{year}{2013}).

\bibitem[{\citenamefont{Shiozaki and Sato}(2014)}]{PhysRevB.90.165114}
\bibinfo{author}{\bibfnamefont{K.}~\bibnamefont{Shiozaki}} \bibnamefont{and}
  \bibinfo{author}{\bibfnamefont{M.}~\bibnamefont{Sato}},
  \bibinfo{journal}{Phys. Rev. B} \textbf{\bibinfo{volume}{90}},
  \bibinfo{pages}{165114} (\bibinfo{year}{2014}).

\bibitem[{\citenamefont{Smith et~al.}(2000)\citenamefont{Smith, Padilla, Vier,
  Nemat-Nasser, and Schultz}}]{PhysRevLett.84.4184}
\bibinfo{author}{\bibfnamefont{D.~R.} \bibnamefont{Smith}},
  \bibinfo{author}{\bibfnamefont{W.~J.} \bibnamefont{Padilla}},
  \bibinfo{author}{\bibfnamefont{D.~C.} \bibnamefont{Vier}},
  \bibinfo{author}{\bibfnamefont{S.~C.} \bibnamefont{Nemat-Nasser}},
  \bibnamefont{and} \bibinfo{author}{\bibfnamefont{S.}~\bibnamefont{Schultz}},
  \bibinfo{journal}{Phys. Rev. Lett.} \textbf{\bibinfo{volume}{84}},
  \bibinfo{pages}{4184} (\bibinfo{year}{2000}).

\bibitem[{\citenamefont{Kohmoto et~al.}(1992)\citenamefont{Kohmoto, Halperin,
  and Wu}}]{PhysRevB.45.13488}
\bibinfo{author}{\bibfnamefont{M.}~\bibnamefont{Kohmoto}},
  \bibinfo{author}{\bibfnamefont{B.~I.} \bibnamefont{Halperin}},
  \bibnamefont{and} \bibinfo{author}{\bibfnamefont{Y.-S.} \bibnamefont{Wu}},
  \bibinfo{journal}{Phys. Rev. B} \textbf{\bibinfo{volume}{45}},
  \bibinfo{pages}{13488} (\bibinfo{year}{1992}).

\bibitem[{\citenamefont{Koshino et~al.}(2001)\citenamefont{Koshino, Aoki,
  Kuroki, Kagoshima, and Osada}}]{PhysRevLett.86.1062}
\bibinfo{author}{\bibfnamefont{M.}~\bibnamefont{Koshino}},
  \bibinfo{author}{\bibfnamefont{H.}~\bibnamefont{Aoki}},
  \bibinfo{author}{\bibfnamefont{K.}~\bibnamefont{Kuroki}},
  \bibinfo{author}{\bibfnamefont{S.}~\bibnamefont{Kagoshima}},
  \bibnamefont{and} \bibinfo{author}{\bibfnamefont{T.}~\bibnamefont{Osada}},
  \bibinfo{journal}{Phys. Rev. Lett.} \textbf{\bibinfo{volume}{86}},
  \bibinfo{pages}{1062} (\bibinfo{year}{2001}).

\end{thebibliography}

\end{document}